\documentclass[sigconf, 9pt]{acmart}

\renewcommand\footnotetextcopyrightpermission[1]{}

\AtBeginDocument{%
  }

\usepackage{amsmath}
\usepackage{subcaption}
\usepackage{algorithmic}
\usepackage{graphicx}
\usepackage{textcomp}
\usepackage{xcolor}
\usepackage{booktabs} 
\usepackage{soul}
\graphicspath{{figs/}}
\usepackage{braket}
\usepackage{etoolbox}
\usepackage{fancyhdr}
\usepackage{multirow}

\setcopyright{acmcopyright}
\copyrightyear{2023}
\acmYear{2023}
\acmDOI{XXXXXXX.XXXXXXX}

\acmConference[Conference acronym 'XX]{Make sure to enter the correct
  conference title from your rights confirmation emai}{June 03--05,
  2018}{Woodstock, NY}
\acmPrice{15.00}
\acmISBN{978-1-4503-XXXX-X/18/06}

\begin{document}

\fancyhead{}
\title{Security Concerns in Quantum Machine Learning as a Service}

\author{Satwik Kundu}
\affiliation{%
  \institution{The Pennsylvania State University}
  \city{University Park}
  \state{PA}
  \country{USA}
  \postcode{16801}}
\email{sxk6259@psu.edu}

\author{Swaroop Ghosh}
\affiliation{%
  \institution{The Pennsylvania State University}
  \city{University Park}
  \state{PA}
  \country{USA}
  \postcode{16801}}
\email{szg212@psu.edu}


\begin{abstract}
Quantum machine learning (QML) is a category of algorithms that employ variational quantum circuits (VQCs) to tackle machine learning tasks. Recent discoveries have shown that QML models can effectively generalize from limited training data samples. This capability has sparked increased interest in deploying these models to address practical, real-world challenges, resulting in the emergence of Quantum Machine Learning as a Service (QMLaaS). QMLaaS represents a hybrid model that utilizes both classical and quantum computing resources. Classical computers play a crucial role in this setup, handling initial pre-processing and subsequent post-processing of data to compensate for the current limitations of quantum hardware. Since this is a new area, very little work exists to paint the whole picture of QMLaaS in the context of known security threats in the domain of classical and quantum machine learning. This SoK paper is aimed to bridge this gap by outlining the complete QMLaaS workflow, which encompasses both the training and inference phases and highlighting significant security concerns involving untrusted classical or quantum providers. QML models contain several sensitive assets, such as the model architecture, training/testing data, encoding techniques, and trained parameters. Unauthorized access to these components could compromise the model's integrity and lead to intellectual property (IP) theft. We pinpoint the critical security issues that must be considered to pave the way for a secure QMLaaS deployment. 

\end{abstract}




\maketitle

\section{Introduction}
Quantum computing is rapidly progressing, with companies like Atom Computing and IBM recently unveiling the largest quantum processors ever developed, boasting 1,225 and 1,121 qubits, respectively \cite{computing2023quantum, gambetta2023hardware}. The significant interest in quantum computing among academic and research communities stems from its potential to offer substantial computational speedups over classical computers for certain problems. Researchers have already begun leveraging these noisy intermediate-scale quantum (NISQ) machines to demonstrate practical utility in this pre-fault-tolerant era \cite{kim2023evidence}. Within this emergent field, quantum machine learning (QML) has also gained considerable attention, merging the power of quantum computing with classical machine learning algorithms. QML heuristically explores the potential of improving learning algorithms by leveraging the unique capabilities of quantum computers, opening new horizons in computational speed and capability. Several QML models have been explored, including quantum support vector machines (QSVMs) \cite{rebentrost2014quantum}, quantum generative adversarial networks (QGANs) \cite{dallaire2018quantum}, and quantum convolutional neural networks (QCNNs) \cite{cong2019quantum}. However, quantum neural networks (QNNs) \cite{schuld2014quest, abbas2021power, farhi2018classification, guerreschi2017practical, schuld2020circuit} stand out as the most notable development, mirroring the structure and function of classical neural networks within a quantum framework.

Training QML models effectively requires integration of both quantum and classical computing resources. Currently, Noisy Intermediate Scale Quantum (NISQ) devices are limited by factors such as qubit count, noise levels, fidelity, and quantum volume. For instance, a quantum computer with 100 qubits is unlikely to reliably run a 100-qubit QML circuit due to inherent noise limitations. To mitigate these limitations, classical techniques are often employed at the outset to preprocess and reduce the size of input data (images or features). This preprocessing ensures that the QML circuit can execute more reliably on the quantum hardware to perform the necessary computations. Furthermore, during the QML training process, although there are quantum-native techniques available for calculating gradients of the parameters, such as the parameter-shift rule \cite{schuld2019evaluating} and simultaneous perturbation stochastic optimization (SPSA) \cite{wiedmann2023empirical, spall1997one}, the task of final parameter optimization still relies on classical optimizers. This reliance is primarily due to the complexities and challenges associated with implementing optimization routines directly on quantum computers. Additionally, once the quantum circuit has processed the data, the outputs generally require further classical processing. This may involve additional computational layers or post-processing steps executed on classical computers to render the quantum computation outputs useful and interpretable. 

With the rise of quantum computing access predominantly provided through the cloud by various startups and companies, the transition to hosting quantum circuits, including QML models, over the cloud, referred to as QMLaaS (Quantum Machine Learning as a Service) is imminent. For QMLaaS to function effectively, it is crucial to ensure secure and efficient communication between classical and quantum resources. However, this interdependence exposes the QMLaaS framework to increased risks of adversarial threats from both the classical and quantum domains\cite{liao2021robust, gong2024enhancing, west2023towards, kundu2022security}.
Untrusted classical cloud providers can jeopardize various assets such as raw training/testing data and the final outputs of QML models, potentially leading to adversarial attacks like model inversion and inference attacks. Similarly, untrusted quantum cloud providers may threaten quantum-specific assets, including the QML architecture and novel state preparation circuit. They could also reroute the QML model's execution to compromised/low-quality hardware, undermining the confidentiality, integrity, and availability of these models. Consequently, it is imperative to conduct thorough studies to assess these security vulnerabilities and develop innovative techniques to ensure the efficient and secure operation of future QMLaaS providers.


\textbf{Why QML Models are at Risk?}
Apart from the vulnerabilities of hybrid QMLaaS, QML models in general face significant security risks due to the following reasons: 

\textbf{High Training Cost:} Currently, accessing quantum computers is significantly more expensive than using classical GPUs. For instance, IBM charges \$1.60 per second to access their superconducting QPUs \cite{ibm_quantum}, which is at least 2,300 times costlier than high-performance GPUs, priced at approximately \$0.0007 per second \cite{gcloud}. AWS Quantum also offers access to a variety of QPUs from providers like IonQ, Rigetti, and IQM, where charges are based on both the number of tasks and the shots used \cite{aws_quantum}. QML models require hundreds of training epochs, each involving thousands of quantum circuit executions, depending on factors like the size of the training dataset, gradient calculation methods, etc. Each circuit execution involves thousands of trials to obtain expectation values, making the training and even partial training of QML models very expensive. While current state-of-the-art machine learning models, such as Gemini \cite{reid2024gemini}, require millions to billions of dollars for training, scaling QML models could potentially cost orders of magnitude more, thereby making them extremely valuable.

\textbf{High Training Time:} Current state-of-the-art ML models, such as GPT-4 \cite{achiam2023gpt} took $\sim 4+$ months for training using thousands of dedicated GPUs. In contrast, quantum resources are both scarce and in high demand. This scarcity leads to long wait times for both hardware access and simulators, whose computation time scales exponentially with the number of qubits. Even users with dedicated access, such as those in Quantum Hubs with a limited number of participants, experience these delays. Consequently, training a large QML model could take a significant amount of time—potentially months to years—due to these extended queue times and the need to execute hundreds of thousands of quantum circuits.

\textbf{Hosting QMLs on the Quantum Cloud:} Since QML providers may not possess their own quantum hardware, they may rely on a third-party quantum cloud for hosting the model. This will lead to the rise of QMLaaS \cite{kundu2024evaluating} providing access to clients only through input-output queries via external APIs. The quantum cloud provider, having white-box access to both the quantum circuit and the expensive training data, could potentially expose these assets to various attacks and thefts \cite{patel2023toward, ma2022qenclave, trochatos2024dynamic, upadhyay2022robust, ayanzadeh2023enigma}. 

\textbf{Miscellaneous intellectual property (IP):} QML models possess various forms of IP. The untrained IPs of a QML model comprise its fundamental architecture—including aspects such as entanglement strategies, the number of parameters, the layer depth, and the measurement basis. Additionally, the training data is often incorporated directly into the state preparation circuit. Trained QML IPs consist of the optimized parameters, which have been fine-tuned through training processes. These parameters, along with the input data used during inference, are also embedded within the state preparation circuit.

In this study, we first provide a comprehensive description of an exemplary hybrid QMLaaS framework. We discuss in detail each stage involved in the training and inference processes of a QML model within a cloud-based environment. Following this, we explore the various security vulnerabilities that could compromise the confidentiality, integrity, and availability of the hybrid QMLaaS framework. Addressing these vulnerabilities is essential for ensuring the secure and efficient operation of QMLaaS. 


\section{Background} \label{background}

\subsection{Quantum Neural Network (QNN)} 
QNN mainly consists of three building blocks: (i) a classical to quantum data encoding (or embedding) circuit, (ii) a parameterized quantum circuit (PQC) whose parameters can be tuned (mostly by an optimizer) to perform the desired task, and (iii) measurement operations. There are a number of different encoding techniques available (basis encoding, amplitude encoding, etc.) but for continuous variables, the most widely used encoding scheme is angle encoding where a variable input classical feature is encoded as a rotation of a qubit along the desired axis \cite{abbas2021power}. As the states produced by a qubit rotation along any axis will repeat in 2$\pi$ intervals, features are generally scaled within 0 to 2$\pi$ (or -$\pi$ to $\pi$) in a data pre-processing step. In this study, we consider $RZ$ gates to encode classical features into their quantum states.

\begin{figure}[!t]
        \vspace{0mm}
        \centering 
        \includegraphics[width=\linewidth]{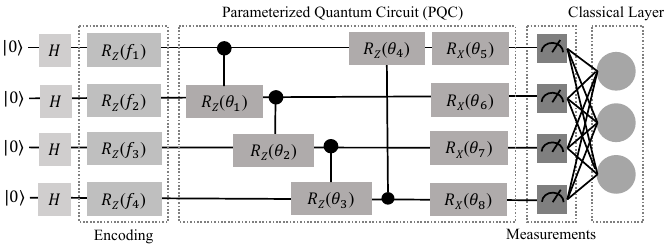}
        \vspace{-4mm}
        \caption{Architecture of a 4-qubit hybrid QNN. Classical features are encoded as angles of quantum rotation gates ($R_Z$). PQC transforms encoded states to explore the search space and entangle features. Measured expectation values are then fed into a classical linear layer for final prediction.}
        \label{qnn_circuit}
        \vspace{-4mm}
\end{figure}

A PQC consists of a sequence of quantum gates whose parameters can be varied to solve a given problem. In QNN, the PQC is the primary and only trainable block to recognize patterns in data. The PQC is composed of entangling operations and parameterized single-qubit rotations. The entanglement operations are a set of multi-qubit operations (that may or may not be parameterized) performed between all of the qubits to generate correlated states and the parametric single-qubit operations are used to search the solution space. 
Finally, the measurement operation causes the qubit state to collapse to either `0' or `1'. We used the expectation value of Pauli-Z to determine the average state of the qubits. 
The measured values are then fed into a classical neuron layer (the number of neurons is equal to the number of classes in the dataset) in our hybrid QNN architecture as shown in Fig. \ref{qnn_circuit}, which performs the final classification task. Other QML architectures may directly apply a softmax function to the measured qubit values or pass them through multiple classical layers for further processing.

\subsection{Quantum Cloud Services}
Recently, there has been a significant increase in the number of companies offering cloud access to quantum hardware. IBM, which employs superconducting transmon qubits for their quantum processing units (QPUs), has recently removed their lower-qubit devices from cloud access \cite{ibm_quantum}. They now offer hardware ranging from 127-qubit to 156-qubit systems, with an error rate per layer of gates as low as 0.6\%. Rigetti, also using superconducting qubits for their quantum processors, has recently started providing cloud access to their 84-qubit Ankaa-2 quantum hardware \cite{rigetti}. This system is known for its higher coherence times and fidelities. Oxford Quantum Circuits (OQC) offers access to up to 32-qubit quantum hardware, which operates on superconducting qubits within a coaxmon architecture \cite{oqc}. Their OQC Toshiko Gen 1 boasts over 96\% 2-qubit gate fidelity and is recognized as the world’s first enterprise-ready platform. IQM provides access to their 20-qubit IQM Radiance, which is expected to be upgradeable to 150-qubit configurations in the near future \cite{iqm}. This platform also utilizes superconducting qubits. QuEra’s Aquila was the first and remains the only publicly accessible 256-qubit neutral atom quantum computer \cite{wurtz2023aquila}. It is based on programmable arrays of neutral rubidium atoms, trapped in a vacuum by tightly focused laser beams. Xanadu, known for developing the  popular PennyLane framework, offers cloud access to their X-Series devices \cite{xanadu}. These are the first photonic quantum computers deployed to the cloud. IonQ provides cloud access to their trapped-ion quantum computers, which achieve 2-qubit fidelity of up to 99.6\% \cite{ionq}. 

\begin{figure*}[!t]
        \vspace{-4mm}
        \centering 
        \includegraphics[width=\textwidth]{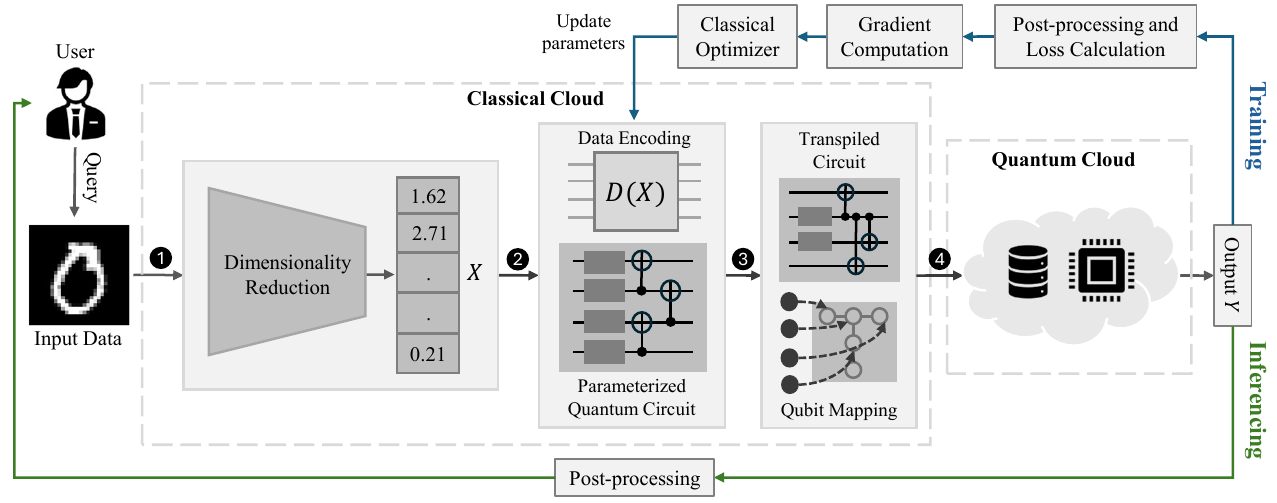}
        \vspace{-6mm}
        \caption{QMLaaS Workflow: (1) Input data is pre-processed using dimensionality reduction (e.g., PCA, autoencoders) and normalized for effective QML training. (2) The reduced features are encoded into a quantum circuit, and a suitable PQC is selected. (3) The circuit is transpiled to match the quantum hardware's topology and basis gates. (4) The circuit is sent to a quantum cloud provider for execution. Training: Post-processing of measured outputs, loss calculation, and parameter updates are performed using a classical optimizer. Inferencing: Outputs are post-processed to return the final vector/label to the user.}
        \label{qmlaas}
        \vspace{-2mm}
\end{figure*}

\section{Training in QMLaaS}
Fig. \ref{qmlaas} presents a detailed workflow of QMLaaS. Training QML models on cloud-based quantum hardware involves a multi-step process, combining classical and quantum computing techniques. The methodology consists of several key stages, integrating classical pre-processing and QNN processing, forming a hybrid system optimized for machine learning tasks. 

\subsection{(Step-1) Data Pre-Processing}
Due to the qubit limitations of current quantum hardware, raw training data must first be pre-processed in a classical cloud to effectively train QML models on large datasets. Thus, upon receiving the input data, the first stage of processing occurs in the classical cloud, encompassing the following steps: 
\begin{itemize}
   \item \textit{Dimensionality Reduction:} This involves reducing the input data dimension to match the qubit capacity of the quantum hardware. For image classification tasks, this reduction may involve resizing or applying dimensionality reduction techniques such as Principal Component Analysis (PCA) \cite{wang2022quantumnat}, Linear Discriminant Analysis (LDA), etc. to extract essential features for classification. Non-linear dimensionaltiy reduction techniques like the convolutional autoencoder (CAE) \cite{alam2021quantum} can also be used which has been found to outperform PCA especially for image classification tasks using hybrid QNNs.
   \item \textit{Normalization:} Next, these extracted features must then be normalized before training. Normalization is crucial because, during the encoding step, features are often passed as rotation angles of quantum gates, unnormalized values can cause features of different classes to appear identical to the QNN if their values differ by multiples of \(2\pi\). There exists several normalization techniques which can be used but few of the most widely used techniques are min-max scaling and max absolute scaling.
\end{itemize}  
The output is a reduced and normalized data matrix \( X \), where each element represents a feature of the input data.

\subsection{(Step-2) Design and Encode} 
The reduced and normalized data \( X \) is then encoded into a quantum-compatible format using a data encoding circuit \( D(X) \). This circuit transforms classical data into quantum states through various techniques such as angle encoding and amplitude encoding. Another critical component of QML circuit design involves selecting an optimal PQC, which serves as the primary trainable block of QML models. Ideally, PQCs with a higher number of parametric gates should enhance performance due to their increased expressive power. However, in practical scenarios, increasing the number of gates, including both 1 and 2-qubit gates, often leads to a higher error rate. This is primarily due to factors such as increased decoherence, gate errors, and the necessity for swap gate insertions. Consequently, employing a brute force approach to select the correct QML circuit for specific quantum hardware may not be effective. In order to address these challenges, recent advancements have introduced efficient quantum circuit search frameworks \cite{anagolum2024elivagar, wang2022quantumnas, du2022quantum, huang2022robust}. These frameworks are designed to identify the most performant circuit for a given quantum hardware setup. The search techniques employed are noise-guided, topology-aware, and data embedding-aware, which have collectively been shown to enable superior performance over traditional circuit design methods. Once the optimal circuit design is identified, it is sent to the cloud provider for execution.

\subsection{(Step-3) Transpile and Map} 
Upon receiving the circuit, it undergoes transpilation and logical to physical qubit mapping to match the basis gates and topology of the cloud quantum hardware. This crucial step ensures the circuit can be efficiently executed on the specific quantum system in use. The transpilation process involves several detailed steps to optimize the quantum circuit and ensure its compatibility with the hardware. These steps include \cite{qiskit_transpiler_docs}:
\begin{itemize}
    \item \textit{Virtual Circuit Optimization:} Simplifying the circuit at a virtual level before mapping it to physical qubits.
    \item \textit{Decomposition of 3+ Qubit Gates:} Breaking down more complex multi-qubit gates into simpler 1- and 2-qubit operations.
    \item \textit{Placement on Physical Qubits:} Assigning logical qubits from the virtual circuit to the physical qubits available on the hardware.
    \item \textit{Routing on Restricted Topology:} Adjusting the circuit pathways to fit the specific qubit connectivity of the hardware.
    \item \textit{Translation to Basis Gates:} Converting the circuit's gates into the set of basic operations supported by the quantum processor.
    \item \textit{Physical Circuit Optimization:} Further refining the circuit to minimize errors and enhance performance after placement and routing.
\end{itemize}
It is important to note that all these processes are performed on a classical computer. This transpilation can be either automated and completed beforehand by the user or handled by the quantum cloud provider. If done beforehand, the transpiled circuit is directly sent for execution; otherwise, it is transpiled by the provider before execution. This ensures that the quantum computations are set up for optimal performance when executed on the cloud quantum hardware.   

\subsection{(Step-4) Execute and Measure} 
Finally, the transpiled quantum circuit is sent to the quantum cloud provider for execution on the chosen quantum hardware. Typically, the circuit execution job is added to a queue for the public quantum hardware, as a large number of users are utilizing the hardware. Once the job reaches the front of the queue, it is executed on the quantum hardware and the required qubits are measured. There are a variety of measurement techniques used to measure qubits like the basis measurement, Pauli measurement (X, Y, Z), quantum state tomography etc. The raw measured values obtained are then subjected to post-processing (classical), which is essential for their practical usage. Depending on the architecture of the Hybrid QML model, this post-processing might include operations like the softmax function, which normalizes the output probabilities, or even integration with classical layers, such as linear layers. This integration is crucial for calculating the loss needed for the training and optimization process.

\subsection{(Step-5) Gradient Calculation} 
To update the parameters of the QML model effectively, the gradient of the parameters needs to be calculated. Unlike classical neural networks, backpropagation is not feasible on quantum computers due to the No-Cloning Theorem \cite{wootters1982single, buvzek1996quantum}, which prohibits copying intermediate quantum states for use in a backward pass. As a result, alternative techniques such as the parameter-shift rule \cite{mitarai2018quantum, schuld2019evaluating} and finite differences are employed to calculate gradients on quantum hardware. The parameter-shift rule, for instance, is quite resource-intensive. To calculate the gradients of \( n \) parameters, it requires \( 2n \) circuit executions. Recognizing the computational demands of such methods, researchers have increasingly turned to a more efficient approach known as Simultaneous Perturbation Stochastic Approximation (SPSA) \cite{spall1998overview, spall1998implementation}. SPSA is a zero-order gradient estimation technique that significantly reduces computational overhead; it requires only 2 circuit executions to estimate the gradients of all parameters, regardless of their number. However, while SPSA offers a dramatic reduction in the number of required circuit executions, it comes with a trade-off: the gradients it produces are noisier compared to those obtained through methods like the parameter shift.

\subsection{(Step-6) Parameter Optimization} 
After calculating the gradients and determining the loss function, a classical optimizer is employed to update the circuit parameters. This optimization step aims to minimize the loss function, thereby improving the model's performance on the training data. Empirically, it has been found that optimizers like Adam and AMSGrad work well in tandem with SPSA for optimizing quantum circuits executed in a noisy environment \cite{wiedmann2023empirical}. Steps 1-5 are iteratively executed until the QML model achieves the required accuracy or a predefined loss threshold is met.

\section{Inferencing in QMLaaS}

\subsection{Hosting QML in Quantum-Classical Cloud}
In the QMLaaS framework, the deployment and inferencing process leverages both classical and quantum computing resources to efficiently process and analyze data. To host a trained QML model on the cloud and provide access via an API, first, a suitable classical cloud platform, such as AWS, Google Cloud, or Azure, and a quantum cloud platform like IBM Quantum or OQC, should be selected, ensuring they support both machine learning and quantum circuit deployments. The trained dimensionality reduction model and the QML circuit need to be packaged in a format compatible with the chosen platforms. Next, a cloud instance should be set up, or a managed service like AWS Lambda, Google Cloud Functions, or Azure Functions can be used to deploy these models. Once deployed, an API endpoint, created using frameworks like Flask or cloud-specific services such as AWS API Gateway, will manage incoming requests. The API processes the data through classical pre-processing, encodes the pre-processed data into the trained QML model, and then performs transpilation to optimize the quantum circuit for specific quantum hardware based on a pre-defined algorithm before sending it to the quantum cloud provider. Finally, the circuit is executed on the designated quantum cloud hardware. Security measures, including API keys or OAuth, should be implemented to protect the API, along with logging and monitoring to ensure smooth and secure operation.
\subsection{Inference Operation}
The workflow initiates when a user submits a query accompanied by input data, such as an image of a handwritten digit. Initially, the data undergoes classical pre-processing in the cloud, which includes dimensionality reduction using techniques like PCA or t-SNE, and normalization to scale the data suitably for quantum processing. The pre-processed data is then encoded into QML model using the encoding technique used while training. Once encoded, the data is sent to the quantum cloud where the trained QML model is executed on available quantum hardware, such as IBM Quantum Experience or Rigetti Aspen. 

Post-execution, the quantum results are sent back to the classical cloud for post-processing. This includes transforming the raw quantum outputs, often probability distributions or measurements, into meaningful classical information through techniques like softmax or linear transformations. The final processed results are then delivered to the user, providing relevant outputs such as classifications or predictions based on the original query. 

\section{Security Concerns}
\subsection{Assets in QMLaaS}
\subsubsection{Training/Testing Data}
Data used for training QML models or during inferencing are critical assets because they are often highly sensitive, difficult to obtain, and expensive to acquire and process \cite{jain2020overview}. In a hybrid QMLaaS framework, this data may be processed locally or over the cloud for tasks like dimensionality reduction, making it vulnerable to threats such as data theft attacks \cite{shokri2015privacy, mireshghallah2020privacy}. Sensitive data, including personal health or financial records, must be handled securely to prevent privacy breaches and legal complications. Moreover, acquiring high-quality data is particularly challenging in specialized domains, requiring significant effort, time, and adherence to regulatory standards. The cost associated with collecting, labeling, and preparing this data further adds to its value, as it directly influences the performance and reliability of machine learning models, making it a crucial and protected resource.


\subsubsection{Data Encoding Circuit}
The data encoding circuit in a QML model is used for embedding classical data into its corresponding quantum state, making it one of the essential components of any QML model. Selecting the optimal data encoding circuit is a challenging task, as it directly influences the performance of variational QML models \cite{li2022concentration, larose2020robust, schuld2021effect, balewski2024quantum}. This process often requires extensive evaluation of different encoding strategies on noisy quantum hardware to identify the most suitable circuit for a given system. Techniques such as Quantum Circuit Search (QCS) \cite{anagolum2024elivagar} have also been employed to optimize the choice of encoding circuits for QML models. The process is further complicated when encoding sensitive or expensive private data, which adds significant value to the data encoding component of the QML model. As a result, if an adversary gains access to this circuit, it could pose a serious threat to the confidentiality and integrity of the QML model.

\subsubsection{PQC Architecture}
During training, the parameters of the quantum gates within the PQC are iteratively optimized to minimize a loss function, enabling the model to perform its designated task. Designing an optimal PQC is a complex and resource-intensive process, as it requires careful consideration of factors such as expressibility, entanglement capability, and the reachability of the quantum states \cite{cerezo2021variational, bharti2022noisy, sim2019expressibility}. Additionally, the PQC must be tailored to the specific quantum hardware, accounting for noise levels, available basis gates, and the device's topology \cite{wang2022quantumnas, anagolum2024elivagar}. However, designing an effective PQC goes beyond these considerations, especially given the challenge of the barren plateau problem, which can hinder optimization when the PQC is too deep, highly entangled, or overly expressive \cite{pesah2021absence, mcclean2018barren, larocca2024review}. Interestingly, recent studies suggest that reducing entanglement in PQCs can actually improve performance, making the design of a robust PQC even more nuanced \cite{bowles2024better, kundu2023wepro}. Furthermore, the intermediate parameter values during training and final parameter values during inferencing can be considered as assets since they take significant time and cost to obtain. Given the significant time, resources, and expertise required to develop a well-functioning PQC, it is considered a valuable asset in the realm of QML.



\subsection{Adversary Motivation}
\begin{figure}[!t]
        \vspace{-0mm}
        \centering 
        \includegraphics[width=\linewidth]{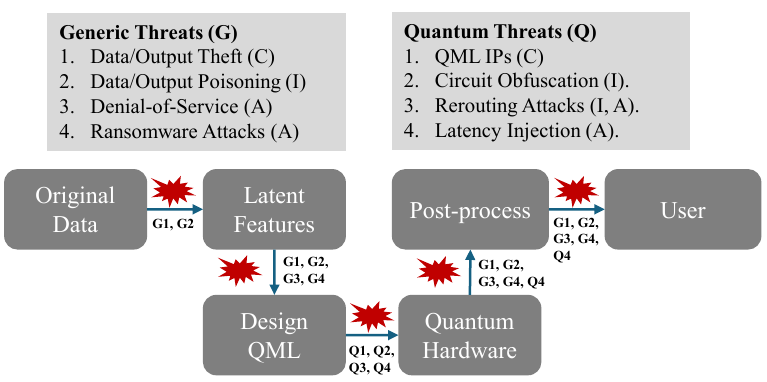}
        \vspace{-6mm}
        \caption{Key Threats to Confidentiality (C), Integrity (I), and Availability (A) in the QMLaaS Pipeline.}
        \label{security}
        \vspace{-4mm}
\end{figure}

Adversary will be motivated to steal the QNN and/or its assets to avoid paying for (i) the time and resources needed to design a QNN from scratch, (ii) the training/inferencing data and (iii) the time and resources needed for training the model. Thus, even though QMLaaS provider easier access to wider variety of quantum hardwares and architectures, it also opens up several security vulnerabilities. In the following section we will discuss few of the major security concerns (Fig. \ref{security}) which comes with QMLaaS and how it affects the confidentiality, integrity and availability of the QML models.

\subsection{Confidentiality}

\subsubsection{Threats from Classical Cloud}
The hybrid QMLaaS framework faces security of data and QML models. Similar to traditional Machine Learning as a Service (MLaaS), there is a risk of raw data theft, either during training or inferencing \cite{shokri2015privacy}. This risk pertains to sensitive data sent for preprocessing on classical cloud systems by untrusted providers especially during the training stage when both data and labels are sent to train the feature extraction model. Even during inference, although the classical cloud may only have access to input data and not the labels, adversaries can employ techniques such as clustering to reverse engineer the original labels or collude with the cloud provider responsible for loss calculation to obtain the labels. Stealing sensitive training data can provide adversaries with confidential business insights and personally identifiable information, which can be exploited for financial gain or competitive advantages. 

\subsubsection{Threats from Quantum Cloud}
In QMLaaS, this threat extends to quantum-encoded data, which becomes vulnerable when handled by untrusted quantum cloud providers. An adversary could potentially extract the encoding circuit from QML models and use it to either sell the circuit or train their own QML model, thereby offering a similar service \cite{wang2024pristiq}. However, training their own QML model would either require collusion with cloud provider responsible for loss calculation or further analysis like majority voting, to extract the labels, as the adversary would not have direct access to them \cite{upadhyay2024quantum}. 
This dual threat underscores the unique security risks associated with both the classical and quantum components of the QMLaaS framework.


Furthermore, QML models that incorporate novel encoding techniques and architectures are particularly vulnerable to threats from potentially untrusted quantum cloud-based adversaries. Given that these cloud providers would have white-box access to the QML circuits, there is a heightened risk of intellectual property (IP) theft. Such access enables adversaries to steal these specialized architectures and encoding techniques, which they could then potentially sell to competing businesses. Similarly, trained QML models hosted on these untrusted platforms are also at risk of being stolen, showcasing the confidentiality and security challenges faced in the QMLaaS framework. This situation underscores the critical need for robust security measures to protect against the theft of both data and intellectual assets \cite{kundu2024stiq, wang2023qumos, lu2024quantum}.

QML circuits in a cloud-based quantum computing environment are also threatened by physical attacks, especially as users lack direct control over the hardware. As quantum computers increasingly handle sensitive IP through complex algorithms, the risk of these circuits being compromised grows. Malicious insiders in data centers could execute power-based side-channel attacks to extract information about the control pulses used in quantum operations \cite{xu2023exploration}. By analyzing these pulses, attackers can reverse-engineer the gate-level description of the QML circuits, revealing the underlying algorithms or sensitive data embedded within the circuits. Such attacks can compromise the confidentiality of proprietary quantum algorithms and data.

QML circuits are also at significant risk due to state leakage, particularly arising from noisy and erroneous reset operations necessary between circuit executions \cite{xu2024thorough}. When these operations are flawed, residual quantum information from previous executions can persist and carry over to subsequent ones, leading to "horizontal" leakage. This leakage allows attackers to infer sensitive quantum states used in a victim's QML circuits. Additionally, "vertical" leakage, occurring simultaneously between qubits due to issues like crosstalk, further compromises confidentiality by allowing adversaries to extract information from multiple qubits within the same execution.


Finally, even when trained QML models are hosted on trusted cloud providers, they remain susceptible to external threats, such as model stealing or model inversion attacks. Consider a scenario where the cloud-hosted QML model operates as a black box—users do not have access to information about the model's architecture or the dataset on which it was trained, only the formats of the input and output data. In this setup, an adversary could systematically query the cloud-hosted model, gathering significant information that could be used to replicate the model's functionality or extract details about the training data \cite{kundu2024evaluating, fu2024quantumleak, fredrikson2015model}. 

\subsection{Integrity}

\subsubsection{Threats from Classical Cloud}
The raw training/ testing data sent to the classical cloud for dimensionality reduction is also vulnerable to threats such as data poisoning attacks \cite{wang2022threats, nelson2008exploiting, zhao2017efficient, geiping2020witches, feng2019learning}. Adversary could either tamper or introduce adversarial examples to the original raw data. Such manipulations can corrupt the data before it is even encoded into quantum format, undermining the QML model's reliability from the outset. 

\subsubsection{Threats from Quantum Cloud}
The integrity of QML models is particularly at risk in the hybrid cloud-based QMLaaS framework due to a variety of factors that extend beyond classical data integrity issues. These include challenges specific to quantum data and quantum circuits. Especially, once the data is encoded into quantum circuits, it faces additional threats. Adversarial obfuscation of the quantum encoded circuit can severely impair the model's performance. This issue is particularly critical for the primary trainable block of the QML model, the PQC. Adversaries may attempt to manipulate the circuit architecture or even tamper the parameters or the measurement outputs, leading to significant performance degradation \cite{upadhyay2024stealthy, upadhyay2024obfuscating, upadhyay2022robust, upadhyay6trustworthy}.

Furthermore, an adversary could exploit crosstalk between qubits to launch a fault injection attack on a victim's QML circuit execution in multi-tenant computing environment \cite{ash2020analysis}. By continuously operating their own qubits with quantum gates, such as CNOT, the adversary can induce errors in the neighboring qubits used by the victim. This interference degrades the accuracy and reliability of the victim's computational results due to crosstalk. Attackers can also compromise the integrity of QML models by targeting various components within the quantum computing system, including the QPU, Quantum Computer Controller, and Classical Co-processor \cite{xu2023classification}. By manipulating physical qubits and couplings through voltage changes or electromagnetic radiation, attackers can induce faults that alter quantum operations. They can also interfere with the analog control pulses, modifying their frequency, phase, or envelope to induce errors in gate operations. Furthermore, attacks on the digital specifications used to generate these pulses, as well as classical registers that store critical data, can lead to incorrect quantum operations and corrupted outputs. In the Classical Co-processor, similar attacks on classical registers can distort the computations and optimizations in QML, ultimately compromising the model's accuracy and reliability.


Another concern arises from the variety of available quantum hardware. An adversary operating within the cloud infrastructure might allocate lower-quality quantum hardware for executing QML circuits \cite{phalak2021quantum}. This allocation strategy might be motivated by the lower costs associated with running circuits on less capable hardware. However, this not only compromises the integrity of the QML models but also detrimentally affects their performance. Each of these factors underscores the complex integrity challenges faced by QML models in a cloud-based, hybrid quantum-classical environment, necessitating robust strategies to ensure the security and reliability of these systems.
\subsection{Availability}
\subsubsection{Threats from Classical Cloud}
When pre-/post-processing data over classical cloud, several security concerns related to availability can rise as well. For instance, denial of service (DoS) attack could be targeted at the classical computing resources, effectively disrupting training/inferencing process and rendering computational resources inaccessible \cite{mahjabin2017survey, bonguet2017survey}. Another critical concern is ransomware attacks where malicious adversaries can encrypt sensitive pre-/post-processed data or computational resources, demanding payment for access restoration. Such attacks not only halt model training but could also result in loss of valuable data \cite{brewer2016ransomware, thamer2021survey}. 

\subsubsection{Threats from Quantum Cloud}
Adversaries based in the quantum cloud could also launch DoS attacks, disrupting the availability of quantum hardware, or they could withhold execution outputs from quantum hardware, leading to ransomware attacks. Furthermore, as discussed earlier, the QMLaaS workflow involves using both classical and quantum resources during the training and inference stages. This dependency creates a potential for latency, which can be exploited by a cloud-based adversary to delay the training and inference processes. During the training phase, an adversary might reroute quantum circuit executions to slower or more congested quantum hardware. This intentional rerouting to cheaper hardware with longer queue times can substantially delay the gradient calculation process, thereby increasing the overall training time. Similar tactics can be applied during the inference stage, where timely responses are often critical. Adversaries could introduce latency to degrade the performance of the QML models, affecting the responsiveness of the service. Additionally, adversaries might induce artificial demand on specific quantum hardware, creating a bottleneck. This can be achieved by submitting a large number of low-priority tasks to certain quantum processors, causing genuine QML tasks to experience significant delays \cite{phalak2024qualiti}. These tactics can greatly affect the availability of QMLaaS, compromising the efficiency and reliability of the service. 

\section{Conclusion}
The rapid development of quantum computers and the growing interest in harnessing their practical utility have sparked significant exploration in various applications, with quantum machine learning (QML) being one of the most intensively researched domains. The implementation of QML models is expected to lead to the emergence of QMLaaS, a hybrid framework that leverages both classical and quantum resources to deliver QML services. This study has provided a detailed description of each component of the QMLaaS framework and highlighted the various security concerns inherent in this hybrid approach. Addressing these security issues will be crucial for achieving a secure and reliable QMLaaS deployment in the future.

\bibliographystyle{unsrt}
\bibliography{refs}

\end{document}